\documentclass[preprint,superscriptaddress,showpacs,12pt,a4paper,oneside,UTF8,nofootinbib]{revtex4}
\usepackage{amssymb}
\usepackage[utf8]{inputenc}
\usepackage[T1]{fontenc}
\usepackage[colorlinks=true,linkcolor=black,citecolor=blue,urlcolor=blue]{hyperref}
\usepackage{longtable}
\usepackage{graphicx}    
\usepackage{bm}          
\usepackage{dcolumn}    
\usepackage{amssymb}
\usepackage{color}
\usepackage{CJK}
\usepackage{url}
\usepackage{booktabs}
\usepackage{amsmath,amssymb,mathrsfs,amsfonts,nicefrac}
\usepackage{mathtools}

\begin{document}

\title{Collective stimulated Brillouin scattering modes of two crossing laser beams with shared ion acoustic wave}
\author{Jie Qiu}
\affiliation{Institute of Applied Physics and Computational
Mathematics, Beijing, 100094, China}
\author{Liang Hao \footnote{Corresponding author: hao\_liang@iapcm.ac.cn}}
\affiliation{Institute of Applied Physics and Computational
Mathematics, Beijing, 100094, China}
\author{Lihua Cao}
\affiliation{Institute of Applied Physics and Computational
Mathematics, Beijing, 100094, China} \affiliation{Key Laboratory of
HEDP of the Ministry of Education, CAPT, Peking University, Beijing,
100871, China}
\author{Shiyang Zou}
\affiliation{Institute of Applied Physics and Computational
Mathematics, Beijing, 100094, China}


\begin{abstract}
The overlapping of multiple beams is common in inertial confinement
fusion (ICF), making the collective stimulated Brillouin scattering
(SBS) with shared ion acoustic wave (IAW) potentially important
because of the effectively larger laser intensities to drive the
instability. In this work, based on a linear kinetic model, an exact
analytic solution for the convective amplification of SBS with the
shared IAW modes stimulated by two overlapped beams is presented.
From this solution, effects of the wavelength difference, crossing angle, polarization states,
and finite beam overlapping volume of the two laser beams on the shared
IAW modes are studied. It is found that a wavelength difference of
several nanometers between the laser beams has negligible effects,
except for a very small crossing angle about one degree.
However, the crossing angle, beam polarization states,
and finite beam overlapping volume can have significant
influences on the
shared IAW modes. Furthermore, the out-of-plane modes, in which the
wavevectors of daughter waves lie in the different planes from the
two overlapped beams, is found to be important for certain
polarization states and crossing angles of the laser beams with the
finite beam overlapping volume. This work is helpful to comprehend
and estimate the collective SBS with shared IAW in ICF experiments.
\end{abstract}


\maketitle


\section{Introduction}
In inertial confinement fusion (ICF), the overlapping of multiple
beams is common for both indirect-drive and direct-drive
schemes~\cite{Myatt2014MultiBeamLPI,Kirkwood2013LPIReview}. This
leads to complex multibeam laser-plasma instabilities. One important
example is crossed-beam energy transfer (CBET) between two
beams~\cite{Michel2009CbetTuneSymmetry,Michel2010CbetNIF,Michel2012StochaIonHeating,Moody2012CBetPowerTransfer,Hao2016Cbet3D},
which can redistribute laser energy, alter drive symmetry, and
modify hydrodynamic conditions. Apart from CBET, the collective
stimulated Brillouin scattering (SBS) and the collective stimulated
Raman scattering (SRS) of multiple beams with shared daughter waves
can also be important, since the temporal growth rate and convective
gain are expected to depend on the combined laser intensities.
Experimentally, it is observed that the energy amplification factor
increases with the number of pump beams and significant scattered
light losses are produced in novel backward directions due to the
collective
coupling~\cite{Seka2002MultiBeamSBS,Kirkwood2011MultiBeamBackScatter,Michel2015MultiBeamSRSICF,Neuville2016CollectiveSBS,Depierreux2019MBeamSRSSBS}.
Thus the laser-target coupling is significantly impaired.
Understanding, and in some cases mitigating these endemic processes, is essential for optimizing ICF implosions.

The shared daughter wave can be a common Langmuir/ion acoustic wave
or scattered light wave, termed as the shared plasma wave (SP) mode
and shared scattered light wave (SL) mode, respectively.
Theoretically, the homogeneous temporal growth rate for collective
SP and SL modes of multiple beams has been investigated in
a general framework under fluid
description~\cite{DuBois1992OverlapLaser,Xiao2019LinearMultiBeam,Zhao2021MultiBeamSRSPolyChrome}.
However, for most practical cases in ICF, SRS and SBS instabilities
are spatial
problems~\cite{Forslund1975SBS-SRSAnaly,Hao2014SRSSBSScatter,Hao2019LPIOuterRingSGIII,Ji2021SRSRescatter},
for which theoretical work for the collective modes is lacking.
Although two-dimensional (2D) particle-in-cell simulations
have been performed recently to study SRS of two overlapped
laser
beams~\cite{Yang2020SRSTwoBeamGrowth,Zhao2021MultiBeamSRSPolyChrome},
the out-of-plane modes, where the wavevectors of daughter waves lie
in the different planes from the plane of two overlapped beams,
have not been considered, since two-dimensional simulations
restrict all wavevectors in the same plane. In this work,
theoretical study for the SP modes of collective SBS in convective
regime is performed. By a linear kinetic model, an exact
analytic solution for convective amplification of SP modes of two overlapped beams
is obtained in the limit of strong damping of plasma waves and
negligible pump depletion. Based on this solution, impacts of the
wavelength difference, crossing angle, polarization states, and
finite beam overlapping volume of the two laser beams on the
collective SBS modes with shared ion acoustic wave (IAW) are studied
systematically. Especially, the out-of-plane modes are discussed,
which might be important in ICF experiments. This work is helpful to
comprehend and estimate the collective SBS
with shared IAW in ICF experiments. 

This paper is organized as follows: In Section~\ref{sec:model}, the
theoretical model for SP modes is presented, where an analytic
solution for its convective amplification is given. In
Section~\ref{sec:SBS}, impacts of the wavelength difference,
crossing angle, polarization states, and finite beam overlapping volume of two laser beams on the
scattered wavelength and spatial amplification of the collective SBS
modes with shared IAW are investigated, and the importance of
out-of-plane modes relative to in-plane modes are also discussed. In
Section~\ref{sec:conc}, the conclusions as well as some discussions
are given.

\section{Theoretical model for SP modes of two beams}
\label{sec:model}

\subsection{Matching geometry of the SP modes}
The SRS or SBS stimulated by a single laser beam is a
resonant three-wave parametric instability process where the
incident wave decays into a plasma wave and a scattered wave. It
occurs when the plasma wave is resonant with the ponderomotive force
created by beating of the laser wave and scattered light wave, and the
scattered wave is resonant with the transverse current created by
beating between the laser wave and the density perturbation of the
plasma wave. Thus, the phase matching condition is required,
\begin{equation}
  \mathbf{K}_0=\mathbf{K}_s+\mathbf{K}_{\rm es}
  \label{eq:phasematchfour}
\end{equation}
where $\mathbf{K}_i\equiv(\omega_i/c,\mathbf{k}_i)$ with subscripts
$i=\rm 0,s,es$ are the four-wavevectors (comprising the wave
frequency $\omega_i$ and wavenumber vector $\mathbf{k}$) of the
laser beam ($i=0$), scattered wave ($i=\rm s$) and the plasma wave
($i=\rm es$) respectively.

When there are two overlapped beams at four-wavevectors
$\mathbf{K}_{01}$ and $\mathbf{K}_{02}$, the beating of these two
waves with a common plasma wave at four-wavevector $\mathbf{K}_{\rm
es}$ would generate scattered waves at
$\mathbf{K}_{s_1}=\mathbf{K}_{01}-\mathbf{K}_{\rm es}$ and
$\mathbf{K}_{s_2}=\mathbf{K}_{02}-\mathbf{K}_{\rm es}$. (The
anti-Strokes components are ignored here.) The coupling of these
generated scattered waves with the laser beams ($\mathbf{K}_{s_1}$
with $\mathbf{K}_{02}$ or $\mathbf{K}_{s_2}$ with $\mathbf{K}_{01}$)
generates two new plasma waves at $\mathbf{K}_{\rm es}\pm \Delta
\mathbf{K}_0$, where $\Delta \mathbf{K}_0\equiv
\mathbf{K}_{01}-\mathbf{K}_{02}$. Again, the coupling of these two
plasma waves with the pump beams generates two new scattered wave at
$\mathbf{K}_{s_1}+\Delta \mathbf{K}_0$ and
$\mathbf{K}_{s_2}-\Delta\mathbf{K}_0$, etc. Finally, a series of
scattered waves at $\mathbf{K}_{s_1}+n\Delta\mathbf{K}_0$ and
$\mathbf{K}_{s_2}-n\Delta\mathbf{K}_0$, and plasma waves at
$\mathbf{K}_{\rm es}\pm n\Delta\mathbf{K}_0$ ($n=0,1\cdots$) can be
generated. However, only those plasma waves and scattered waves with
four-wavevectors nearly resonant with the natural modes are
important. Assuming that $\mathbf{K}_{\rm es}$, $\mathbf{K}_{s_1}$
and $\mathbf{K}_{s_2}$ are resonant with the natural modes, then the
waves with $n\geq 1$ can be ignored when the two laser beams are
incoherent in the sense that either $\Delta \omega_0$ or $|\Delta
\mathbf{k}_0|$ is large enough to exceed the resonant peak widths of
the natural modes of plasma and scattered waves. In such cases, each
laser beam develops its scattered wave independently, and there are
five coupled waves remained, including two laser waves and their
corresponding scattered waves, and one common plasma wave. The
five-wave matching condition can be written as
\begin{equation}
  \mathbf{K}_{0\alpha}=\mathbf{K}_{s_\alpha}+\mathbf{K}_{\rm es},
  \label{eq:KmatchSP}
\end{equation}
where the subscripts $_{0\alpha}$, $_{\rm s_\alpha}$
($\alpha=1,2$) and $_{\rm es}$ represent the incident wave, scattered
wave and plasma wave, respectively.

\begin{figure}[!h]
\centering
\includegraphics[angle=0,width=0.65\textwidth]{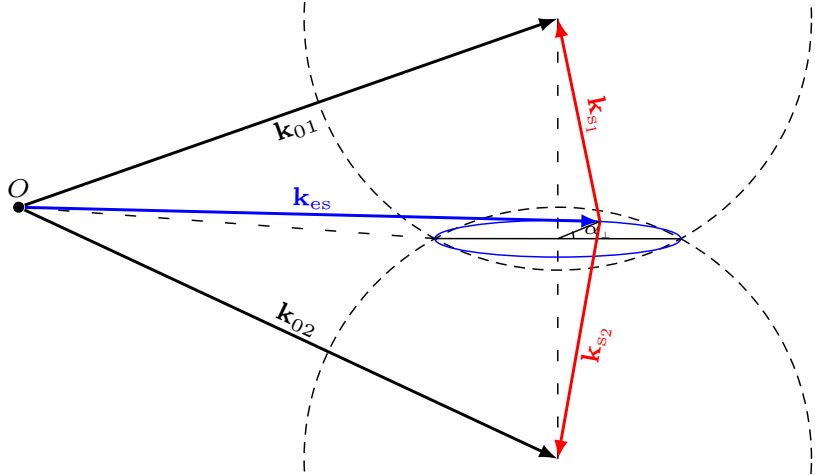}
\caption{The sharing of one common plasma wave by  two overlapping beams (the general case with nonzero wavelength difference is demonstrated).
The incident wave, scattered wave and the plasma wave are shown in
black, red, and blue, respectively.
The wavevectors of the all possible shared plasma waves are located on a circle (in blue) in a plane perpendicular to $\mathbf{k}_{01}-\mathbf{k}_{02}$.
The out-of-plane angle $-\pi<\alpha_\perp\leq \pi$ is defined as the angle between the
$(\mathbf{k}_{01},\mathbf{k}_{02})$-plane and the
$(\mathbf{k}_{s_1},\mathbf{k}_{s_2})$-plane.
}
  \label{Fig:SharePWGeo}
\end{figure}

The matching condition
(\ref{eq:KmatchSP}), together with the dispersion relation of the
electromagnetic waves (EMWs),
\begin{equation}
\omega_{s_\alpha}^2=\omega_{\rm pe}^2+c^2\mathbf{k}_{s_\alpha}^2,
  \label{eq:EMWDispersion}
\end{equation}
can determine $\mathbf{k}_{\rm es}$ as a function of
$\mathbf{k}_{0\alpha}$, $\omega_{0\alpha}$, $\omega_{\rm es}$, and
the out-of-plane angle $\alpha_\perp$ defined as the angle between the
$(\mathbf{k}_{01},\mathbf{k}_{02})$-plane and the
$(\mathbf{k}_{s_1},\mathbf{k}_{s_2})$-plane.
Intuitively, the five-wave coupling geometry of SP modes is
shown in Fig.~\ref{Fig:SharePWGeo}, where
the two spheres formed by
$\mathbf{k}_{s_1}$ and $\mathbf{k}_{s_2}$ have radii
$k_{s_\alpha}=\sqrt{(\omega_{0\alpha}-\omega_{\rm es})^2-\omega_{\rm pe}^2}/c$ ($\alpha=1,2$) according to the matching
condition (\ref{eq:KmatchSP}) and the dispersion relation (\ref{eq:EMWDispersion}),
and
$\mathbf{k}_{\rm es}$ is
determined by the intersection of these two spheres,
which generally defines a circle in a plane
perpendicular to $\Delta\mathbf{k}_0$.
Different SP modes (corresponding to different points on the $\mathbf{k}_{\rm es}$-circle)
can be parameterized by the out-of-plane angle $-\pi<\alpha_\perp\leq \pi$, where
the in-plane modes, for which $\mathbf{k}_{\rm es}$, $\mathbf{k}_{s_1}$ and $\mathbf{k}_{s_2}$ are coplanar with $\mathbf{k}_{01}$ and $\mathbf{k}_{02}$, have $\alpha_\perp=0$ or $\alpha_\perp=\pi$.

Along this line, the wavevector of the shared plasma wave of three
overlapped beams is determined by intersection between the three
spheres formed by $\mathbf{k}_{s_\alpha}$ ($\alpha=1,2,3$), which defines two points when existing.
For more than three beams, a common
plasma wave exists only when the intersection points between the
three spheres formed by $\mathbf{k}_{s_\alpha}$ ($\alpha=1,2,3$)
happen to be located on spheres formed by other
$\mathbf{k}_{s_\alpha}$ ($\alpha=4,\cdots$). This imposes a
stringent requirement on the beam symmetry as noticed by \cite{Xiao2019LinearMultiBeam}.
For an indirect drive,
the required beam symmetry can be
satisfied only in a quite limited volume near the laser entrance
hole (LEH), when the incident laser beams are angularly distributed in a highly symmetric
configuration~\cite{Myatt2014MultiBeamLPI}.
Beyond this region, not only the beam symmetry condition is hard to be satisfied, but
also the overlapping volume of multiple beams drops rapidly.
Consequently, compared to multiple beam overlapping, effects of two
beam overlapping would become more important.

\subsection{Convective amplification of the SP modes}
The equation for the EMW is
\begin{equation}
   \partial_t^2\mathbf{A}+c^2\nabla\times (\nabla \times \mathbf{A})=c^2\mu_0
   \mathbf{J},
   \label{eq:EMWA0}
\end{equation}
where $\mathbf{A}$ is the potential vector and $\mathbf{J}$ is the
transverse current.
Using the cold-fluid approximation for the transverse electron
motion $m_e\mathbf{v}_e=e\mathbf{A}$, the transverse current
$\mathbf{J}=-en_e\mathbf{v}_e=-e^2n_e\mathbf{A}/m_e$,
where $m_e$, $e$, $\mathbf{v}_e$, $n_e$ are the electron mass,
electron charge, electron quiver velocity and electron number density, respectively.
Substituting $\mathbf{J}$ into
Eq.~(\ref{eq:EMWA0}) and taking the normalization
$\boldsymbol{a}=e\mathbf{A}/m_e c$ yield
\begin{equation}
  (\partial_t^2+\omega_{\rm pe}^2)\boldsymbol{a}+c^2\nabla\times (\nabla \times
  \boldsymbol{a})
  =-\omega_{\rm pe}^2\frac{\delta n_{\rm es}}{n_0}\boldsymbol{a},
  \label{eq:EMWGeneral}
\end{equation}
where $n_e=n_0+\delta n_{\rm es}$ is decomposed into
the unperturbed electron density $n_0$ and
the perturbed electron density $\delta n_{\rm es}$,
and $\omega_{\rm pe}=\sqrt{e^2 n_0/\epsilon_0 m_e}$ is the plasma frequency.

In the envelope approximation for five-wave coupling of SP modes,
\begin{equation}
  \boldsymbol{a}(t,\mathbf{r})=\sum_{\alpha=1,2}\frac{1}{2}(\tilde{\boldsymbol{a}}_{0\alpha}e^{j\Psi_{0\alpha}}+cc.)+
  \sum_{\alpha=1,2}\frac{1}{2}(\tilde{\boldsymbol{a}}_{s_\alpha}e^{j\Psi_{s_\alpha}}+cc.),
  \label{eq:EMenvelop}
\end{equation}
and
\begin{equation}
\delta n_{\rm es}(t,\mathbf{r})=\frac{1}{2}(\delta \tilde{n}_{\rm es}e^{j\Psi_{\rm es}}+cc.),
  \label{eq:nesEnvelop}
\end{equation}
where $\tilde{\boldsymbol{a}}_i$ is the complex vector amplitude of EMW,
$\delta \tilde{n}_{es}$ is the complex amplitude of density perturbation,
and the phase $\Psi_i\equiv -j\omega_i t+j\mathbf{k}_i\cdot \mathbf{r}$
with subscript $i={0\alpha}$,$\rm s_\alpha$,$\rm es$ for the laser wave, scattered wave and plasma wave, respectively.
The envelope approximation holds when $|\nabla a_i|/|a_i|\ll k_i$,
$|\partial_t a_i|/|a_i|\ll \omega_i$, $|\nabla n_{\rm es}|/n_{\rm
es}\ll k_{\rm es}$ and $|\partial_t n_{\rm es}|/n_{\rm es}\ll
\omega_{\rm es}$.
Using the phase matching condition (\ref{eq:KmatchSP}),
equations for the electromagnetic components of
$\tilde{\boldsymbol{a}}_i$, which satisfy $\mathbf{k}_i\cdot
\tilde{\boldsymbol{a}}_i=0$, can be derived from
Eqs.~(\ref{eq:EMWGeneral}-\ref{eq:nesEnvelop}) as
\begin{equation}
  \begin{aligned}
  & \mathcal{L}_{\rm em_{\rm s_\alpha}}\boldsymbol{a}_{s_\alpha}=-\frac{j \omega_{\rm pe}^2}{4\omega_{s_\alpha}}\frac{\delta n_{\rm es}^*}{n_0}\mathbf{n}_{s_\alpha}\times(\boldsymbol{a}_{0\alpha}\times \mathbf{n}_{s_\alpha}), \\
  & \mathcal{L}_{\rm em_{0\alpha}}\boldsymbol{a}_{0\alpha}=-\frac{j \omega_{\rm pe}^2}{4\omega_{0\alpha}}\frac{\delta n_{\rm es}}{n_{0}}\mathbf{n}_{0\alpha}\times(\boldsymbol{a}_{s_\alpha}\times
  \mathbf{n}_{0\alpha}).
  \end{aligned}
  \label{eq:EqnEMW}
\end{equation}
For simplicity, the tilde in the complex amplitude is dropped
in Eq.~(\ref{eq:EqnEMW}) and the following paper.
$\mathbf{n}_i=\mathbf{k}_i/k_i$ is defined as an unit vector along
$\mathbf{k}_i$, thus the term $\mathbf{n}_i \times
(\boldsymbol{a}_j\times
\mathbf{n}_i)=\boldsymbol{a}_j-(\mathbf{n}_i\cdot
\boldsymbol{a}_j)\mathbf{n}_i$ is the projection of
$\boldsymbol{a}_j$ onto the plane perpendicular to $\mathbf{k}_i$.
The operator $\mathcal{L}_{\mathrm{em}_i}$ is defined by
\begin{equation}
  \mathcal{L}_{\mathrm{em}_i}\equiv
  \partial_t+\frac{c^2\mathbf{k}}{\omega_i}\cdot \nabla+\frac{c^2}{2\omega_i}\nabla
  \cdot\mathbf{k},
  \label{eq:defLem}
\end{equation}
where 
the $\nabla\cdot \mathbf{k}$ term arises from the plasma
inhomogeneity. 
Then, for a steady-state convective solution in the strong damping regime of homogeneous plasma,
$\mathcal{L}_{\mathrm{em}_i}=c^2\mathbf{k}_{i}\cdot \nabla/\omega_i$.
While the plasma response to the ponderomotive drive has the following expression~\cite{Strozzi2008RayBackScatter},
\begin{equation}
  \frac{\delta n_{\rm es}}{n_0}=-\gamma_{\rm pm} \frac{{k}_{\rm es}^2c^2}{2 \omega_{\rm pe}^2} \sum_{\alpha=1,2}\boldsymbol{a}_{0\alpha}\cdot \boldsymbol{a}_{s_\alpha}^*,
  \label{eq:Eqnes}
\end{equation}
where the ponderomotive response function $\gamma_{\rm pm}$ is
\begin{equation}
  \gamma_{\rm pm}(\omega_{\rm es},k_{\rm es})\equiv \frac{(1+\chi_{I})\chi_{e}}{\epsilon},
\end{equation}
where $\epsilon=1+\chi_I+\chi_e$ is the dielectric function, and
$\chi_I(\omega_{\rm es},k_{\rm es})=\sum_\beta \chi_{\rm
i\beta}(\omega_{\rm es},k_{\rm es})$ and $\chi_e$ are the ion
susceptibility (summed over ion species $\beta$) and electron
susceptibility, respectively.
For simplicity, the flow
velocity is assumed to be zero for all species in the following.
Nevertheless, the above formula can be applied to the non-zero flow
case by replacing $\omega_{\rm es}$ appearing in $\chi_{\rm
i\beta}(\omega_{\rm es},k_{\rm es})$ with $\omega_{\rm
es}-\mathbf{k}_{\rm es}\cdot \mathbf{u}_\beta$ if species $\beta$
were to flow with velocity $\mathbf{u}_\beta$.

Considering the polarization states of the EMWs, the EMW complex vectors can be written as
$\boldsymbol{a}_{s_\alpha}=a_{s_\alpha}\mathbf{e}_{s_\alpha}$ and
$\boldsymbol{a}_{0\alpha}=a_{0\alpha}\mathbf{e}_{0\alpha}$, where
$\mathbf{e}_i$ is a unit vector along the polarization direction.
Using the fact
$\mathbf{n}_{s_\alpha}\cdot \mathbf{e}_{s_\alpha}=\mathbf{n}_{0\alpha}\cdot \mathbf{e}_{0\alpha}=0$,
it can be proven
\begin{equation}
\mathbf{e}_{s_\alpha}\cdot
[\mathbf{n}_{s_\alpha}\times(\mathbf{e}_{0\alpha}\times
\mathbf{n}_{s_\alpha})]=\mathbf{e}_{0\alpha}\cdot
[\mathbf{n}_{0\alpha}\times(\mathbf{e}_{s_\alpha}\times
\mathbf{n}_{0\alpha})]=\mathbf{e}_{s_\alpha}\cdot
\mathbf{e}_{0\alpha}.
  \label{eq:cosphiform1}
\end{equation}
According to Eq.~(\ref{eq:EqnEMW}),
the component of $\boldsymbol{a}_{s_\alpha}$ that can be convectively
amplified is along the direction of
$\mathbf{n}_{s_\alpha}\times(\mathbf{e}_{0\alpha}\times
\mathbf{n}_{s_\alpha})$, so $\mathbf{e}_{s_\alpha}$ is parallel to
$\mathbf{n}_{s_\alpha}\times(\mathbf{e}_{0\alpha}\times
\mathbf{n}_{s_\alpha})$, yielding the polarization alignment factor
\begin{equation}
  \cos\varphi_\alpha\equiv \mathbf{e}_{0\alpha}\cdot \mathbf{e}_{s_\alpha}=|\mathbf{e}_{0\alpha}\times \mathbf{n}_{s_\alpha}|,
  \label{eq:cosphialpSP}
\end{equation}
where $\varphi_\alpha$ is
the angle between $\mathbf{e}_{0\alpha}$ and $\mathbf{e}_{s_\alpha}$.
Then, using Eqs.~(\ref{eq:cosphiform1}-\ref{eq:cosphialpSP}), for the steady-state convective solution,
Eqs.~(\ref{eq:EqnEMW}-\ref{eq:Eqnes}) can be simplified as
\begin{equation}
\begin{aligned}
  &\frac{\delta n_{\rm es}}{n_0}=-\gamma_{\rm pm} \frac{{k}_{\rm es}^2c^2}{2 \omega_{\rm pe}^2}\sum_{\alpha=1,2} \cos\varphi_\alpha a_{0\alpha} a_{s_\alpha}^*, \\
  & \mathbf{k}_{s_\alpha}\cdot \nabla a_{s_\alpha}=-\frac{j \omega_{\rm pe}^2}{4c^2}\frac{\delta n_{\rm es}^*}{n_0}\cos\varphi_\alpha a_{0\alpha}, \\
  & \mathbf{k}_{0\alpha}\cdot \nabla a_{0\alpha}=-\frac{j \omega_{\rm pe}^2}{4c^2}\frac{\delta n_{\rm es}}{n_0}\cos\varphi_\alpha
  a_{s_\alpha},
\end{aligned}
  \label{eq:eqEMWVSP}
\end{equation}
Substituting $\delta n_{\rm es}$ into equations for $a_{s_\alpha}$, we get
\begin{align}
  & \mathbf{n}_{s_1}\cdot \nabla a_{s_1} =\kappa_1(a_{s_1}+r_a a_{s_2}) \label{eq:scatter1}, \\
  & \mathbf{n}_{s_2}\cdot \nabla a_{s_2}
  =\kappa_2(a_{s_2}+a_{s_1}/r_a),
  \label{eq:scatter2}
\end{align}
where $r_a \equiv a_{02}^*\cos\varphi_2/a_{01}^*\cos\varphi_1$,
and the gain coefficient for single beam $\alpha$ is
\begin{equation}
\kappa_\alpha
={\mathrm{Im}[\gamma_{\rm pm}]{k}_{\rm
es}^2|a_{01}|^2\cos^2\varphi_\alpha}/{8k_{s_\alpha}}
  \label{eq:gainSingle}
\end{equation}
Eqs.~(\ref{eq:scatter1}-\ref{eq:scatter2}) are two-dimensional in
nature since $\mathbf{n}_{s_1}$ and $\mathbf{n}_{s_2}$ are along
different directions. They can be solved for a beam
overlapping volume over which the pump depletion of $a_{0\alpha}$ is
negligible.
It is convenient to choose a (in most cases) non-orthogonal
coordinate system $\boldsymbol{x}=x_1\mathbf{n}_{s_1}+x_2\mathbf{n}_{s_2}$, in which
$\mathbf{n}_{s_1}\cdot \nabla=\partial/\partial {x_1}$ and
$\mathbf{n}_{s_2}\cdot \nabla=\partial/\partial{x_2}$.
As derived in Appendix~\ref{app:laplaceSP}, the solution
can be written as
\begin{equation}
  \begin{bmatrix} a_{s_1}({x_1},{x_2}) \\ a_{s_2}({x_1},{x_2}) \end{bmatrix}=
  \int_0^{x_2}\mathbf{G}_1(x_1,x_2-\xi_2)a_{s_1}({x_1}=0,\xi_2)d\xi_2+
  \int_0^{x_1}\mathbf{G}_2(x_1-\xi_1,x_2)a_{s_2}(\xi_1,x_2=0)d\xi_1,
  \label{eq:SPas1as2}
\end{equation}
where
\begin{equation}
  \mathbf{G}_1({x_1},{x_2})=\begin{bmatrix} G_{11} \\ G_{21} \end{bmatrix}=
  \begin{bmatrix}
 \kappa_2\kappa_1x_1e^{\kappa_1{x_1}+\kappa_2{x_2}}\frac{I_1[2\sqrt{\kappa_1\kappa_2{x_1}{x_2}}]}{\sqrt{\kappa_1\kappa_2{x_1}{x_2}}}\\
 (\kappa_2/r_a) e^{\kappa_1{x_1}+\kappa_2{x_2}}I_0[2\sqrt{\kappa_1\kappa_2{x_1}{x_2}}]
  \end{bmatrix}+
  \begin{bmatrix}
e^{\kappa_1{x_1}}\delta({x_2}) \\ 0
\end{bmatrix},~x_1\geq 0,x_2\geq 0
  \label{eq:RespG1}
\end{equation}
is the SP mode response to seed of $a_{s_1}$ at
$\boldsymbol{x}=0$, while
\begin{equation}
  \mathbf{G}_2({x_1},{x_2})= \begin{bmatrix} G_{12} \\ G_{22} \end{bmatrix}=
  \begin{bmatrix}
  r_a\kappa_1 e^{\kappa_1{x_1}+\kappa_2{x_2}}I_0[2\sqrt{\kappa_1\kappa_2{x_1}{x_2}}] \\
  \kappa_1\kappa_2{x_2}e^{\kappa_1{x_1}+\kappa_2{x_2}}\frac{I_1[2\sqrt{\kappa_1\kappa_2{x_1}{x_2}}]}{\sqrt{\kappa_1\kappa_2{x_1}{x_2}}}
  \end{bmatrix}+
  \begin{bmatrix}
0 \\ e^{\kappa_2{x_2}}\delta({x_1})
\end{bmatrix},~x_1\geq 0,x_2\geq 0
  \label{eq:RespG2}
\end{equation}
is the SP mode response to seed of $a_{s_2}$ at
$\boldsymbol{x}=0$, where $I_0$ and $I_1$ are the zero-order and
first-order modified Bessel functions of the first kind,
respectively.

$\mathbf{G}_{1}$ and $\mathbf{G}_2$ comprise two terms. The smaller
second term of $\mathbf{G}_1$ gives us
$a_{s_1}({x_1},{x_2})=e^{\kappa_1{x_1}}a_{s_1}({x_1}=0,x_2)$, and
thus describes the one-dimensional (1D) amplification along
$x_1$-direction of the sidescatter due to beam I alone. This term
can ignite the two-dimensional amplification of the SP modes. For
example, from seed of $a_{s_1}$ at $\boldsymbol{x}=0$, firstly the
sidescatter of beam I generates perturbations $\delta n_{\rm es}$
along the 1D straight line $x_2=0$; this again serves as seeds to
$a_{s_2}$ and amplified along the $\mathbf{n}_{s_2}$ direction,
generating perturbations $\delta n_{\rm es}$ over the 2D
$\boldsymbol{x}$-space; finally the seeding of 2D perturbations of
$\delta n_{\rm es}$ results in the amplification of both $a_{s_1}$
and $a_{s_2}$ over the 2D $\boldsymbol{x}$-space. The steady-state
two-dimensional amplification due to sharing of plasma waves is
described by the first terms of $\mathbf{G}_1$ and $\mathbf{G}_2$,
where the 2D volume $V_{\rm amp}(\boldsymbol{x})\equiv\eta_1x_1\mathbf{n}_1+\eta_2x_2
\mathbf{n}_2$ ($0\leq \eta_{0,1} \leq 1$)
participates in amplifying scattered waves from the seed at $\boldsymbol{x}=0$
 to the point $\boldsymbol{x}=x_1 \mathbf{n}_1+x_2 \mathbf{n}_2$.
 At sufficient gain length $\kappa_1\kappa_2{x_1}{x_2}\gg 1$, for the
first term of $\mathbf{G}_1$, we have
\begin{equation}
  \mathbf{G}_{1} \approx
  \begin{bmatrix}
    \frac{\kappa_2\kappa_1{x_1}}{2\sqrt{\pi}(\kappa_1\kappa_2x_1x_2)^{3/4}}e^{(\sqrt{\kappa_1 {x_1}}+\sqrt{\kappa_2 {x_2}})^2} \\
    \frac{\kappa_2/r_a}{2\sqrt{\pi}(\kappa_1\kappa_2x_1x_2)^{1/4}}e^{(\sqrt{\kappa_1 {x_1}}+\sqrt{\kappa_2
    {x_2}})^2}
  \end{bmatrix}
\end{equation}
Similar results can be obtained for $\mathbf{G}_2$. So all the
response components $G_{11}$, $G_{12}$, $G_{21}$ and $G_{22}$ have
the same dominant asymptotic term $G_A\equiv
e^{(\sqrt{\kappa_1x_1}+\sqrt{\kappa_2x_2})^2}$, which is adopted in the following illustrative analysis instead of the
accurate expression for $\mathbf{G}_{1,2}$.

\begin{figure}[!h]
  \centering
  \includegraphics[angle=0,width=0.95\textwidth]{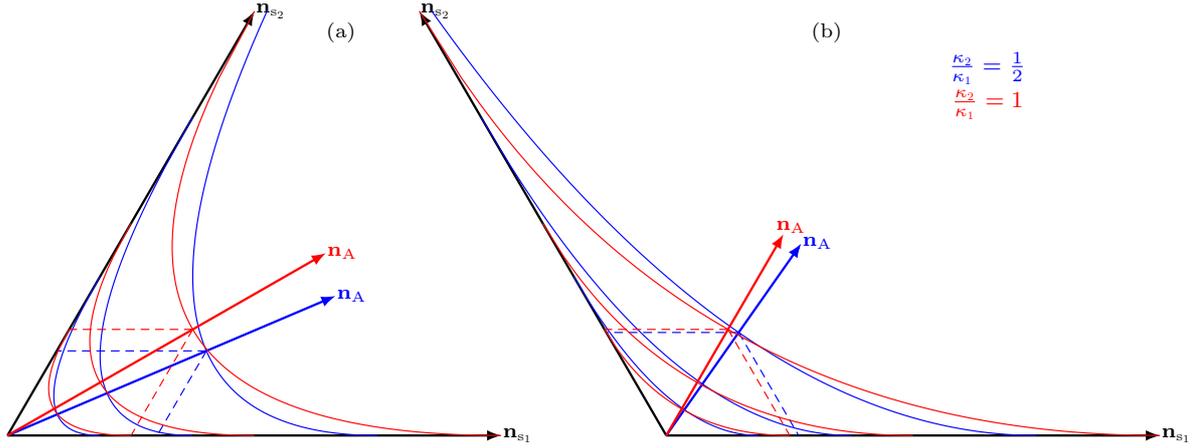}
  \caption{
    The isocurves of asymptotic SP mode response
    $G_A=e^{(\sqrt{\kappa_1x_1}+\sqrt{\kappa_2x_2})^2}$ for (a) $\theta_{12}^s=60^\circ$ and (b) $\theta_{12}^s=120^\circ$,
    when $\kappa_2/\kappa_1=1/2$ (in blue) and $\kappa_2=\kappa_1$ (in red).
    The direction $\mathbf{n}_{\rm A}$ that is parallel to $\nabla G$ is indicated.
    The parallelogram marked by dashed lines is the 2D gain volume that takes parts in amplifying seeds from the origin to the corresponding point.
  }
  \label{Fig:SPRepresentDirection}
\end{figure}

From the asymptotic form $G_A$ of the SP mode response, we can
define a representative gain coefficient to quantify the overall
amplification ability of the SP modes. There is a direction
$\mathbf{n}_A$ parallel to $\nabla G_A$, along which the required
amplification distance is minimum (in the asymptotic sense) for the
seed at a given point (e.g., $\boldsymbol{x}=0$) to achieve a
specific gain, as shown in Fig.~\ref{Fig:SPRepresentDirection}. The
representative gain coefficient $\kappa_{A}\equiv \ln G_A/L_m$,
where $L_m$ is the typical size of the corresponding 2D gain volume
for points along $\mathbf{n}_A$, can be defined as in
Appendix~\ref{sec:SPmeasure}. In our definition, $\kappa_A$ is
approximately of the same magnitude as $\kappa_1+\kappa_2$, where
$\kappa_A=\sqrt{2}(\kappa_1+\kappa_2)$ when $\kappa_1=\kappa_2$,
while $\kappa_A=\kappa_1+\kappa_2$ when $\kappa_1\gg \kappa_2$ or
$\kappa_2\gg \kappa_1$ such that the amplification by one beam
dominates. The ratio $\mathcal{R}_{\rm
A}[\kappa_2/\kappa_1,\theta_{12}^s]\equiv
\kappa_A/(\kappa_1+\kappa_2)$ and the direction of $\mathbf{n}_A$,
can be determined by $\kappa_2/\kappa_1$ and the angle
$\theta^s_{12}$ between $\mathbf{n}_{s_1}$ and $\mathbf{n}_{s_2}$,
as detailed in Appendix~\ref{sec:SPmeasure}.

\section{Analysis of Shared IAW modes of two beams}
\label{sec:SBS}
\begin{figure}[!h]
  \centering
  \includegraphics[angle=0,width=0.5\textwidth]{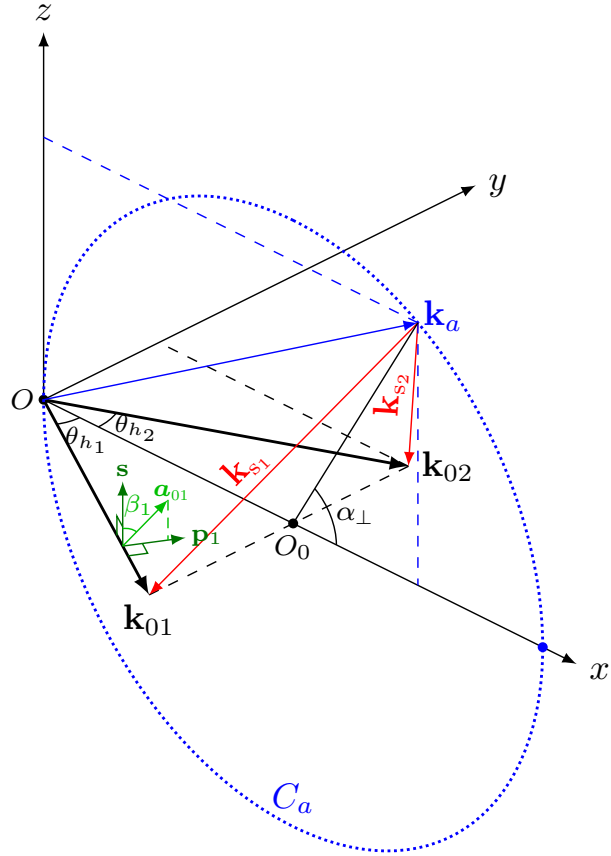}
  \caption{
    The geometry of shared IAW modes of two overlapping beams.
    In the presented coordinate system, xy-plane is chosen to be the $(\mathbf{k}_{01},\mathbf{k}_{02})$-plane with x-axis perpendicular to $\mathbf{k}_{01}-\mathbf{k}_{02}$ and y-axis along $\mathbf{k}_{02}-\mathbf{k}_{01}$, and z-axis is along $\mathbf{k}_{01}\times \mathbf{k}_{02}$.
    The circle $C_a$ (lying in the xz-plane),
    which can be parameterized by the out-of-plane angle $\alpha_\perp$, 
    comprises all possible locations of the endpoints of $\mathbf{k}_a$.
    Beam I or beam II is said to be s-polarized when $\boldsymbol{a}_{0\alpha}$ is along the $\pm\mathbf{s}$ (z-axis) direction,
    and p-polarized when $\boldsymbol{a}_{0\alpha}$ is along the direction of $\pm\mathbf{p}_\alpha=\pm\mathbf{s}\times\mathbf{k}_{0\alpha}$.
    Other linear polarization states of beam I or II are described by the polarization angle $90^\circ\geq\beta_\alpha\geq -90^\circ$,
    which is the angle from $\mathbf{s}$ to $\boldsymbol{a}_{0\alpha}$.
  }
  \label{Fig:SPCoordinate}
\end{figure}

In this section, impacts of the wavelength different, crossing angle
and polarization states of the two laser beams on the collective SBS
modes with shared IAW are investigated. Because $k_{\rm s_1}\approx
k_{01}$ and $k_{\rm s_2}\approx k_{02}$ for SBS, the wave coupling
geometry can be simplified significantly. As shown in
Figure~\ref{Fig:SPCoordinate}, choosing point `O' as the initial
point, the terminal point of $\mathbf{k}_a$ is approximately located
on the circle $C_a$ ($\mathbf{k}_a$-circle), which lies in the plane
perpendicular to $\mathbf{k}_{02}-\mathbf{k}_{01}$, having a center
located at the point `$O_0$' and a radius of the length of $|OO_0|$.
According to definitions of the angles shown in
Fig.~\ref{Fig:SPCoordinate}, the relation
\begin{equation}
  k_{01}\cos \theta_{h_1}=k_{02}\cos\theta_{h_2}=\frac{k_a}{2\cos(\alpha_\perp/2)}
  \label{eq:SPIAWkaeqn}
\end{equation}
can be obtained.
Defining the crossing angle between
$\mathbf{k}_{01}$ and $\mathbf{k}_{02}$ as
$\theta_{12}=\theta_{h_1}+\theta_{h_2}$,
\begin{equation}
  k_a=\frac{2k_{01}k_{02}\sin\theta_{12}\cos(\alpha_\perp/2)}{\sqrt{k_{01}^2+k_{02}^2-2k_{01}k_{02}\cos\theta_{12}}}
  \label{eq:SPIAWka}
\end{equation}
can be derived from Eq.~(\ref{eq:SPIAWkaeqn}).

The representative gain coefficient $\kappa_A$ can be written as
\begin{equation}
\begin{aligned}
\kappa_{A}&=(\kappa_1+\kappa_2) \mathcal{R}_{\rm A}[\kappa_2/\kappa_1,\theta_{12}^s] \\
&= \frac{1}{8}k_{a}^2\mathrm{Im}[\gamma_{\rm pm}]
\left(\frac{|a_{01}|^2}{k_{s_1}}\cos^2\varphi_1+
\frac{|a_{02}|^2}{k_{s_2}}\cos^2\varphi_2\right) \mathcal{R}_{\rm
A}\left[\frac{k_{s_1}|a_{02}|^2\cos^2\varphi_2}{k_{s_2}|a_{01}|^2\cos^2\varphi_1},\theta_{12}\right].
\end{aligned}
\label{eq:kapBASP}
\end{equation}
Here $\theta_{12}^s\approx \theta_{12}$ is taken because
$k_{s_\alpha}\approx k_{0\alpha}$ for SBS, and the polarization
alignment factor $\cos\varphi_\alpha$ as can be calculated from
Eq.~(\ref{eq:cosphialpSP}) depends on the polarization states
($\beta_1,\beta_2$) of the laser beams and geometry
($\alpha_\perp,\theta_{12}$) of the SP mode. For two laser beams
with the same intensity and small wavelength difference,
$|a_0|=|a_{01}|=|a_{02}|$ and $k_s=k_{s_1}\approx k_{s_2}$, the
upper bound of $\kappa_A$ for all possible polarization combinations
of the two laser beams is
\begin{equation}
  \kappa_A^U = \frac{\sqrt{2}k_{a}^2\mathrm{Im}[\gamma_{\rm pm}]|a_{0}|^2}{4k_{s}}.
  \label{eq:maxkapA}
\end{equation}
$\kappa_A^U $ can be achieved when the polarization
direction $\mathbf{e}_{0\alpha}\parallel \mathbf{k}_a\times
\mathbf{k}_{0\alpha}$ ($\alpha=1,2$), which results in
$\mathbf{e}_{0\alpha} \perp \mathbf{k}_{s_\alpha}$ and hence
$\cos\varphi_\alpha=1$, a complete alignment between
$\mathbf{e}_{0\alpha}$ and $\mathbf{e}_{s_\alpha}$.

\begin{figure}[!h]
  \centering
  \includegraphics[angle=0,width=0.95\textwidth]{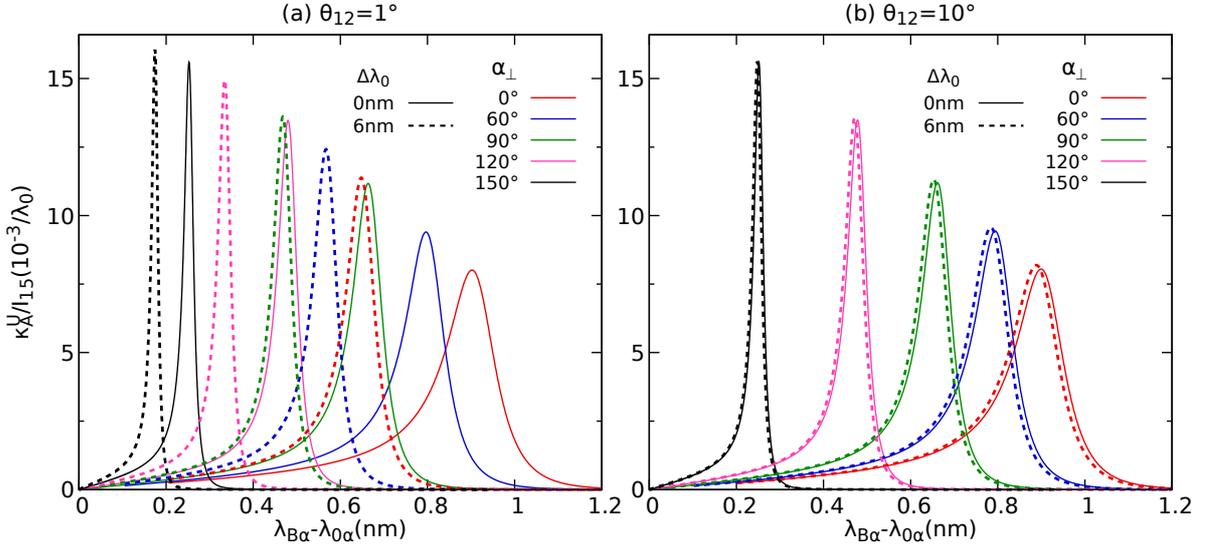}
  \caption{
    $\kappa_{\rm A}^{\rm U}/I_{15}$ versus $\lambda_{B_\alpha}-\lambda_{0\alpha}$ of the shared IAW modes in He plasma
    for two laser beams with wavelength difference $\Delta \lambda_0=0$ and $\Delta \lambda_0=6~\rm nm$, when the beam crossing angles are  (a) $\theta_{12}=1^\circ$ and (b) $\theta_{12}=10^\circ$.
    The two laser beams are at the same intensity ($I_{01}=I_{02}$ and $I_{15}\equiv I_{01}/10^{15}~\rm Wcm^{-2}$) with the vacuum wavelength $\lambda_{01}=351~\rm nm$ and $\lambda_{02}=\lambda_{01}+\Delta \lambda_0$.
    The plasma condition is $n_e=0.06~n_c$, 
    $T_e=2.5~\rm KeV$, $T_e/T_i=3.5$, and zero flow velocity.
  }
  \label{Fig:KB_CmpDwnonzeroHe}
\end{figure}

For two overlapped laser beams with the same vacuum wavelength $\lambda_{01}=\lambda_{02}$,
the $\mathbf{k}_a$-circle is located on the bisecting plane between
$\mathbf{k}_{01}$ and $\mathbf{k}_{02}$ with
$\theta_{h_1}=\theta_{h_2}=\theta_{12}/2$. For a non-zero
wavelength difference $\Delta \lambda_0 \equiv
\lambda_{02}-\lambda_{01} \ll \lambda_{01},\lambda_{02}$,
the plane in which the $\mathbf{k}_a$-circle lies
 deviates from the bisecting plane between $\mathbf{k}_{01}$ and $\mathbf{k}_{02}$, with an angle
\begin{equation}
  \Delta \theta=\frac{\theta_{h_1}-\theta_{h_2}}{2}\approx
\frac{k_{01}-k_{02}}{2k_{01}\tan (\theta_{12}/2)}=\frac{\Delta
\lambda_0}{2\lambda_{01}\tan (\theta_{12}/2)}\frac{1}{1-n_e/n_c}.
\end{equation}
From the expression for $\Delta \theta$, it can be expected that the
effects of the laser wavelength difference on the SP modes are
significant only when $\Delta \lambda_0/\lambda_0$ is comparable to
$2\tan(\theta_{12}/2)$. For $\Delta \lambda_0/\lambda_0 \ll
2\tan(\theta_{12}/2)$, $\Delta \theta$ is quite small, therefore,
$\theta_{h1}\approx \theta_{h2}\approx \theta_{12}/2$ and the
dependence of $\mathbf{k}_a$ on $\Delta\lambda_0$ is rather weak.
Consequently, the effects of the laser wavelength difference on the
SP modes are negligible. As an example, $\kappa_{\rm A}^{\rm U}$
versus $\lambda_{B_\alpha}-\lambda_{0\alpha}$ are shown in
Fig.~\ref{Fig:KB_CmpDwnonzeroHe}(a) and
Fig.~\ref{Fig:KB_CmpDwnonzeroHe}(b) for $\Delta\lambda_0=0~\rm nm$
and $\Delta \lambda_0=6~\rm nm$, respectively, where a typical
plasma condition at LEH~\cite{Michel2013CbertSaturation}
$n_e=0.06~n_c$, $T_e=2.5~\rm KeV$ and $T_e/T_i=3.5$ is chosen. For
simplicity, the flow velocity which only leads a Doppler wavelength
shift is assumed to be zero, and the gain coefficient is normalized
by $I_{15}=I_{01}[~\rm Wcm^{-2}]/10^{15}$. Here, $n_c$ is the
critical density for Beam I. In Fig.~\ref{Fig:KB_CmpDwnonzeroHe}(a),
$\theta_{12}=1^\circ$, so for $\Delta \lambda_0=6~\rm nm$,  $\Delta
\lambda_0/\lambda_0 \approx 2\tan(\theta_{12}/2)$. As expected,
$\kappa_{\rm A}^{\rm U}$ changes significantly when $\Delta
\lambda_0$ varies from zero to $6~\rm nm$. In
Fig.~\ref{Fig:KB_CmpDwnonzeroHe}(b), $\theta_{12}=10^\circ$, so for
$\Delta \lambda_0=6~\rm nm$, $\Delta \lambda_0/\lambda_0<
0.2\tan(\theta_{12}/2)$. As expected, the change of $\kappa_{\rm
A}^{\rm U}$ is negligible when $\Delta \lambda_0$ varies from zero
to $6~\rm nm$. The small crossing angle cases with
$2\tan(\theta_{12}/2)\lesssim \Delta \lambda_0/\lambda_0 $ should be
of interest in the attempt to suppress laser-plasma parametric
instabilities with broadband
lasers~\cite{Zhao2017SRSInCohrLight,Zhao2019SuppressParaInstabMultiFrequencyLight}.
In most cases of ICF, the crossing angle $\theta_{12}$ is typically
not too small, but the laser wavelength difference is much shorter
than the vacuum wavelength~\cite{Michel2011WaveSchemeICF}, making
its effects on the SP modes negligible. Therefore, in the following
analysis, mainly the case of $\Delta \lambda_0=0$ is discussed for
the SP modes.

When $\Delta \lambda_0=0$, 
Eq.~(\ref{eq:SPIAWka}) can be further simplified as
\begin{equation}
  k_a=2k_{01}\cos({\theta_{12}}/{2})\cos({\alpha_\perp}/{2}),
  \label{eq:SPIAWkaEqual}
\end{equation}
and the scattering angle between $k_{s_\alpha}$ and $k_{0\alpha}$ is
\begin{equation}
  \theta_{\rm scat}=\pi-2\arccos[\cos(\theta_{12}/2)\cos(\alpha_\perp/2)].
  \label{eq:SPths}
\end{equation}
In the limiting case $\alpha_\perp \sim 180^\circ$, $\theta_{\rm
scat}\sim 0$ and $\mathbf{k}_{s_\alpha}$ is nearly in the forward
scattering direction of $\mathbf{k}_{0\alpha}$. Such near-forward
SBS needs very long time to develop, and the final saturation stage
usually contains significant contribution from the anti-Strokes
waves~\cite{Corvo1988ForwardSBS,Mckinstrie1997NearForwardSBS}.
Hence, we do not consider these cases here, and restrict
$\theta_{12}\leq 150^\circ$ and $\alpha_\perp\leq 150^\circ$, which
guarantees $\theta_{\rm scat}>7.6^\circ$.

\begin{figure}[!h]
  \centering
  \includegraphics[angle=0,width=0.95\textwidth]{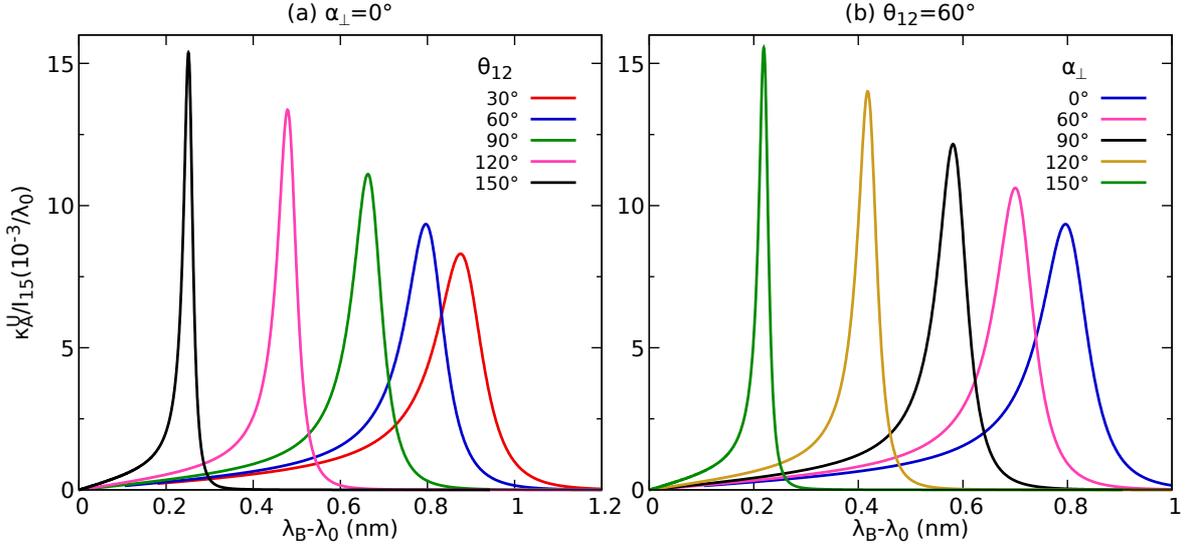}
  \caption{
    $\kappa_{\rm A}^{\rm U}/I_{15}$ versus $\lambda_B-\lambda_0$ of the shared IAW modes in He plasma for
    (a) $\alpha_\perp=0^\circ$ at a variety of $\theta_{12}$
    and (b) $\theta_{12}=60^\circ$ at a variety of $\alpha_\perp$.
    The two laser beams are at the same intensity with the same vacuum wavelength (351~nm).
    The plasma condition is $n_e=0.06~n_c$, $T_e=2.5~\rm KeV$, $T_e/T_i=3.5$, and zero flow velocity.
  }
  \label{Fig:KBSP_HeN}
\end{figure}

Fig.~\ref{Fig:KBSP_HeN} shows $\kappa_{A}^U/I_{15}$ versus
$\lambda_B-\lambda_{0}$ in He plasma for (a) SP modes with different
laser beam crossing angles $\theta_{12}$ and (b) SP modes with
different out-of-plane angles $\alpha_\perp$, when two laser beams
with the same intensity and vacuum wavelength are assumed.
The scattered wavelength at which $\kappa_{A}^U$ peaks decreases with increasing
$\theta_{12}$ or $\alpha_\perp$, while the corresponding peak value
increases with $\theta_{12}$ or $\alpha_\perp$. Since $\kappa_A^U
\propto k_a^2\mathrm{Im}[\gamma_{\rm pm}]$ and $k_a$ decreases with
increasing $\theta_{12}$ or $\alpha_\perp$ from
Eq.~(\ref{eq:SPIAWkaEqual}), these characteristics are mainly due to
properties of $k_a^2\mathrm{Im}[\gamma_{\rm pm}]$, which peaks at
$\omega_a\propto k_a$ and its peak value increases with decreasing
$k_a$, as discussed in Appendix~\ref{app:Gprop}.

%

Now we consider the effects of polarization states of the laser beams on the SP mode
through the dependence of $\kappa_A/\kappa_A^U$ on different polarization combinations of beam I and beam II.
For two laser beams with the same intensity,
\begin{equation}
  \frac{\kappa_A}{\kappa_A^U}=\frac{1}{2\sqrt{2}}(\cos^2\varphi_1+\cos^2\varphi_2)\mathcal{R}_{\rm A}[\cos^2\varphi_2/\cos^2\varphi_1,\theta_{12}]
  \label{eq:kAfacPolar}
\end{equation}
${\kappa_A}/{\kappa_A^U}$ depends only on the geometry
($\alpha_\perp,\theta_{12}$) of the SP mode, and the polarization
alignment between the laser beams and scattered waves. The most
favored SP mode by beam $\alpha$, for which the polarization of
laser beam and the scattered light is in full alignment
($\cos\varphi_\alpha=1$), has $\mathbf{e}_{0\alpha}\perp
\mathbf{k}_a$, as discussed above. Denoting an arbitrary
polarization state of the laser beam by the polarization angle
$\beta_\alpha$ ($-90^\circ<\beta_\alpha\leq 90^\circ$) as shown in
Fig.~\ref{Fig:SPCoordinate}, the most favored out-of-plane angle by
beam I is $\alpha_\perp=-2\arctan[\sin(\theta_{12}/2)\tan\beta_1]$,
while $\alpha_\perp=2\arctan[\sin(\theta_{12}/2)\tan\beta_2]$ is
most favored by beam II. Except for s-polarization
($\beta_\alpha=0$), the SP modes with non-zero out-of-plane angles
are accentuated. When $\beta_1=-\beta_2$, the polarization states of
beam I and beam II are symmetric with respect to the bisecting plane
between $\mathbf{k}_{01}$ and $\mathbf{k}_{02}$, therefore, at one
and the same $\alpha_\perp$, both beam I and beam II are fully
aligned with their respective scattered waves in polarization,
making the value of $\kappa_A/\kappa_A^U$ reach one for this
$\alpha_\perp$. For a nonzero $\beta_1+\beta_2$, the symmetry
between beam I and beam II is broken after taking into account their
polarization states, the most favored out-of-plane angle
$\alpha_\perp$ that results in a full polarization alignment between
beam I and its scattered wave ($\cos\varphi_1=1$) deviates from the
most favored $\alpha_\perp$ by beam II. Consequently, the maximum
value of $\kappa_A/\kappa_A^U$ labeled as
$\max_{\alpha_\perp}[\kappa_A/\kappa_A^U]$ should be less than one.

\begin{figure}[!h]
  \centering
  \includegraphics[angle=0,width=0.95\textwidth]{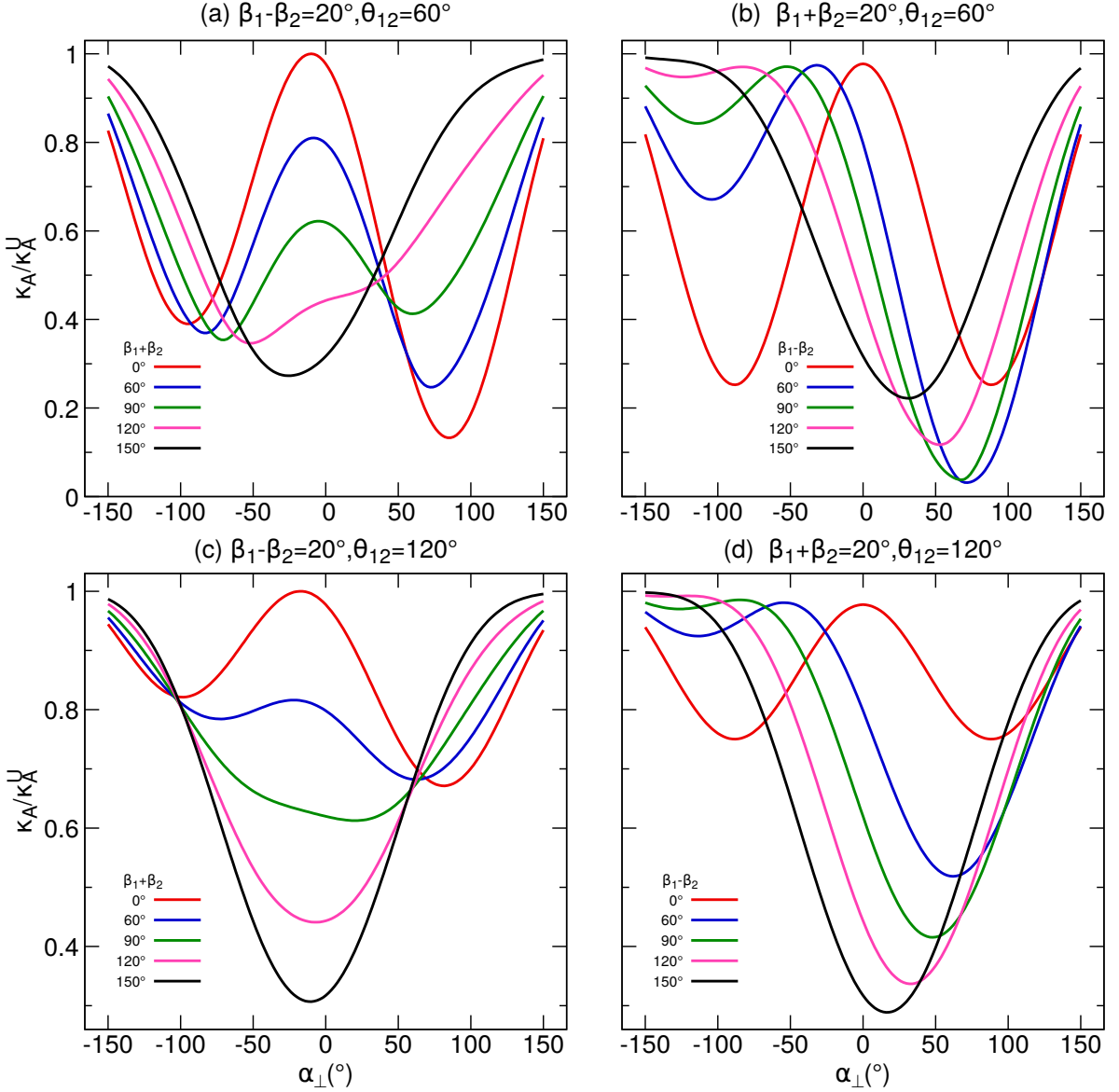}
  \caption{
    $\kappa_A/\kappa_A^U$ versus $\alpha_\perp$ for shared IAW modes of two laser beams with
    crossing angle (a-b) $\theta_{12}=60^\circ$ and (c-d) $\theta_{12}=120^\circ$.
  }
  \label{Fig:SPkapA_thp}
\end{figure}

Fig.~\ref{Fig:SPkapA_thp} shows
the typical variation of $\kappa_A/\kappa_A^U$ with $\alpha_\perp$ for the cases
$\theta_{12}=60^\circ$ and $\theta_{12}=120^\circ$,
where effects of the degree of polarization symmetry breaking (characterized by $\beta_1+\beta_2$) between beam I and II are illustrated in panels (a) and (c),
and the overall effects of deviation from s-polarization of beam I and II (characterized by $\beta_1-\beta_2$)  are illustrated in panels (b) and (d).
One remarkable feature is that
different polarization states can modify the gain coefficient $\kappa_A$
significantly for $|\alpha_\perp|\lesssim 90^\circ$. As a
consequence, the SP mode with some out-of-plane angle, for example,
the angle for which the overlapping geometry permits a long gain length,
can be effectively suppressed by choosing a proper combination of $\beta_1$ and $\beta_2$.
However, for large $|\alpha_\perp|$, the modification by the
polarization states becomes insignificant, because the scattering angles
between $k_{0\alpha}$ and $k_{s_\alpha}$ are small,
and thus
the polarization alignment factors $\cos\varphi_\alpha$ are less sensitive to $\beta_\alpha$.
Furthermore, it can be noticed that
the polarization modification is more significant for an acute crossing angle
$\theta_{12}\leq 90^\circ$ than an obtuse crossing angle
$\theta_{12}>90^\circ$.
In fact, for the former, $\cos\varphi_{1,2}$ can varies from zero to one,
where $\cos\varphi_{1,2}=0$, which corresponds to a complete polarization misalignment between the
laser beams and the scattered waves, occurs when $\beta_1=-\beta_2=\pm
\arcsin[\tan\theta_h]$ and
$\alpha_\perp=\arctan[(\sin\theta_h\tan\beta_1)^{-1}]$;
while for the latter, $\cos\varphi_{1,2}$ can only vary from  $|\cos\theta_{12}|$ to one,
where the minimum of $\cos\varphi_{1,2}$ occurs when both beams are p-polarized
 ($\beta_1=\beta_2=\pm 90^\circ$) and $\alpha_\perp=0$.
Finally, there is a local maximum of $\kappa_A/\kappa_A^U$
in the interior of our considered interval
$-150^\circ<\alpha_\perp<150^\circ$, and at this maximum the value
of $\kappa_A/\kappa_A^{U}$ is mainly determined by $\beta_1+\beta_2$
(the degree of polarization symmetry breaking); while the
corresponding out-of-plane angle $\alpha_{\perp}^M$ is mainly
determined by $\beta_1-\beta_2$ (the overall polarization deviation
from s-polarization). The mode corresponding to the interior local
maximum of $\kappa_A/\kappa_A^U$ is of great interest, since among
the SP modes with relatively small out-of-plane angles and hence
large temporal growth rate  $\gamma_0 \propto \sqrt{k_a}\propto
\sqrt{\cos(\alpha_\perp/2)}$~\cite{Montgomery2016InstabIndirectDriveICF},
it is most vulnerable to be excited.

\begin{figure}[!h]
  \centering
  \includegraphics[angle=0,width=0.97\textwidth]{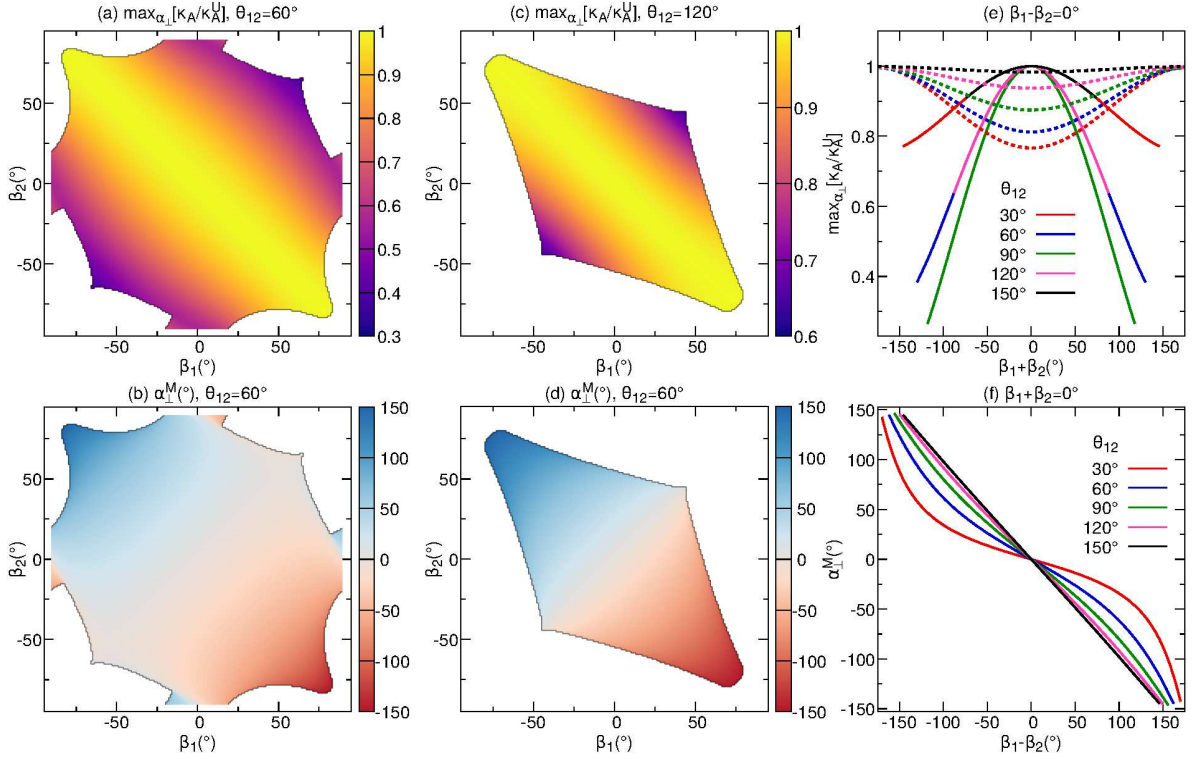}
  \caption{
    (a-d) $\max_{\alpha_\perp}[\kappa_A/\kappa_A^{U}]$ and $\alpha_\perp^M$ versus $\beta_1$ and $\beta_2$ for the interior local maximum of $\kappa_A/\kappa_A^{U}$ when beam crossing angles are $60^\circ$ and $120^\circ$.
    The boundary of the displayed map corresponds to $(\beta_1,\beta_2)$ where the local maximum begins to disappear.
    (e)
    $\max_{\alpha_\perp}[\kappa_A/\kappa_A^{U}]$ versus $\beta_1+\beta_2$ (solid curves)
    inside the interval $-150^\circ<\alpha_\perp<150^\circ$ for different crossing angles, when $\beta_1=\beta_2$ is taken.
    Notice that curves for $\theta_{12}<90^\circ$ and for $180^\circ-\theta_{12}$ coincide except that for the former, the local maximum exists over a larger range of $\beta_1+\beta_2$.
    For comparison,
    the larger one of $\kappa_A/\kappa_A^U$ at the endpoints $\alpha_\perp=\pm 150^\circ$ is also displayed as the dashed curves.
    (f) $\alpha_\perp^M$ versus $\beta_1-\beta_2$, when $\beta_1=-\beta_2$ is taken.
  }
  \label{Fig:SBSSPPolarVar}
\end{figure}

The dependence of the interior local
$\max_{\alpha_\perp}[\kappa_A/\kappa_A^U]$ and its corresponding
$\alpha_{\perp}^M$ on the polarization combination
($\beta_1,\beta_2$) is shown in Figs.~\ref{Fig:SBSSPPolarVar}(a,b)
for case $\theta_{12}=60^\circ$ and
Figs.~\ref{Fig:SBSSPPolarVar}(c,d) for case $\theta_{12}=120^\circ$
respectively. Fig.~\ref{Fig:SBSSPPolarVar}(e) shows the local
$\max_{\alpha_\perp}[\kappa_A/\kappa_A^U]$ versus $\beta_1+\beta_2$
(taking $\beta_1-\beta_2=0$), since the local
$\max_{\alpha_\perp}[\kappa_A/\kappa_A^{U}]$ is mainly determined by
$\beta_1+\beta_2$, while Fig.~\ref{Fig:SBSSPPolarVar}(f) shows
$\alpha_\perp^M$ versus $\beta_1-\beta_2$ (taking
$\beta_1+\beta_2=0$), since $\alpha_{\perp}^M$ is mainly determined
by $\beta_1-\beta_2$. At a small $\beta_1+\beta_2$,
$\kappa_A/\kappa_A^U$ has a local maximum value in the interior of
interval $-150^\circ<\alpha_\perp<150^\circ$ as shown in
Fig.~\ref{Fig:SPkapA_thp}, and this local maximum value can be
larger than $\kappa_A/\kappa_A^U$ at the endpoints $\alpha_\perp=\pm
150^\circ$. With the increase of $\beta_1+\beta_2$, the interior
local maximum value decreases until it disappears at some
$\beta_1+\beta_2$ as shown in Fig.~\ref{Fig:SPkapA_thp}. As a
result, increasing $\beta_1+\beta_2$ to decrease
$\max_{\alpha_\perp}[\kappa_A/\kappa_A^U]$ for relatively small
$|\alpha_\perp|$ can be an effective method to suppress the
collective SBS modes with shared IAW, especially for
$\theta_{12}\sim 90^\circ$. On the other hand, $|\alpha_\perp^M|$
increases from zero towards $180^\circ$ when $|\beta_1-\beta_2|$
increases from zero to $180^\circ$. Therefore, due to modification
of polarization states on $\kappa_A$, depending on $\beta_1-\beta_2$
the most favored out-of-plane angle of the SP modes can deviate
significantly from zero, especially for a large crossing angle. This
implies the potentially important roles of out-of-plane SP modes.

\begin{figure}[!h]
  \centering
  \includegraphics[angle=0,width=0.95\textwidth]{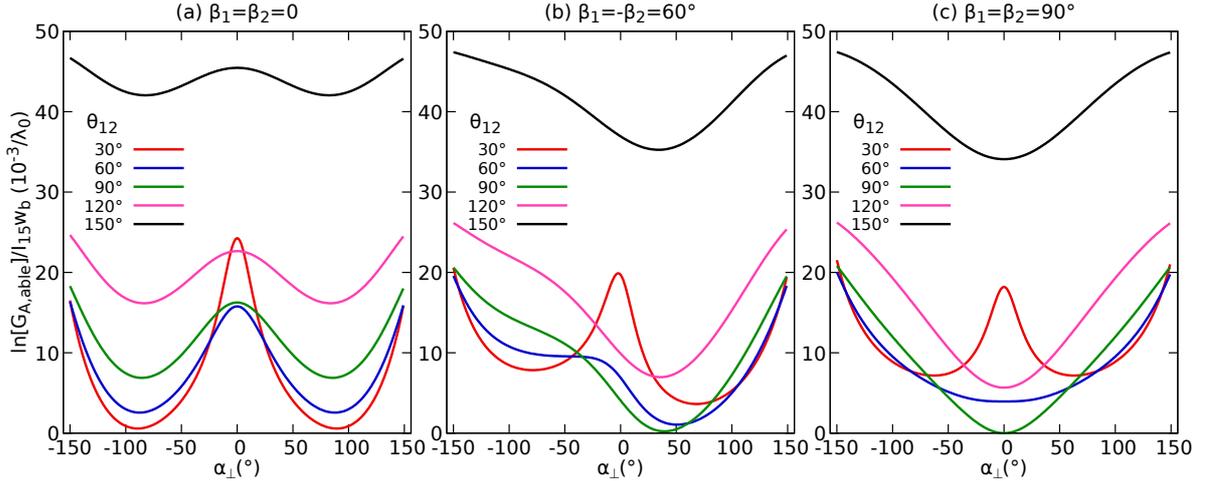}
  \caption{
    $\ln G_{\rm A,able}/I_{15}w_b$ versus $\alpha_\perp$ for (a) $\beta_1=\beta_2=0$, (b) $\beta_1=-\beta_2=60^\circ$ and (c) $\beta_1=\beta_2=90^\circ$.
    The two laser beams are at the same wavelength (351~nm) and intensity ($10^{15}~\rm W/cm^2$).
    The plasma condition $n_e=0.06~n_c$, $T_e=2.5~\rm KeV$, $T_e/T_i=3.5$, and zero flow velocity for He plasma is taken.
  }
  \label{Fig:SP_GainL}
\end{figure}

In practice, the laser beam width is finite, and usually much
smaller than the beam length along the laser propagation direction.
Thus, a finite beam overlapping volume dependent on the width and
crossing angle of the two overlapped beams is formed.
For such case, the parallelogram gain volume $V_{\rm amp}(\boldsymbol{x})=\eta_1x_1\mathbf{n}_{s_1}+\eta_2 x_2 \mathbf{n}_{s_2}$
($0\leq \eta_{1,2} \leq 1$), that takes part in the amplification of
the collective SP modes, must be enclosed in the beam overlapping
volume. Then, the largest achievable asymptotic gain $G_{\rm
A,able}$ for the SP modes is the maximum value of $G_{\rm A}\equiv
e^{(\sqrt{\kappa_1x_1}+\sqrt{\kappa_2x_2})^2}$ subject to
\begin{align}
  \mathrm{Proj}_{\perp \mathbf{k}_{01}}[V_{\rm amp}(\boldsymbol{x})]\leq w_{\rm b},~
  \mathrm{Proj}_{\perp \mathbf{k}_{02}}[V_{\rm amp}(\boldsymbol{x})]\leq w_{\rm b},  \label{eq:GainLcon}
\end{align}
where $w_b$ is the diameter of laser beam at the overlapping point,
$\mathrm{Proj}_{\perp \mathbf{k}_{0\alpha}}[V_{\rm
amp}(\boldsymbol{x})]$ is the longest spatial scale of the
projection of $V_{\rm amp}$ onto a plane perpendicular to
$\mathbf{k}_{0\alpha}$, of which the calculation is given in
Appendix~\ref{app:proj}. The constraint condition
(\ref{eq:GainLcon}) ensures the gain volume can be enclosed within
the beam overlapping region. Taking into account the finite beam
width, the overlapping efficiency between the gain volume and the
beam overlapping volume is highest (complete overlapping for
$x_1=x_2$) when $\alpha_\perp \sim 0,180^\circ$ and decreases when
$|\alpha_\perp|$ becomes closer to $90^\circ$.
However, the decreasing speed becomes slower with the increasing
crossing angle $\theta_{\rm 12}$. It can be estimated that the
overlapping efficiency decreases about 74\%, 50\%, 29\%, 13\%, and
only 3\% when $\alpha_\perp$ changes from zero to $90^\circ$, for
the crossing angles of $30^\circ$, $60^\circ$, $90^\circ$,
$120^\circ$ and $150^\circ$, respectively. In
Fig.~\ref{Fig:SP_GainL}, $\ln G_{\rm A,able}/w_bI_{15}$ versus
$\alpha_\perp$ is shown for three polarization combinations
$\beta_1=\beta_2=0$, $\beta_1=-\beta_2=60^\circ$, and
$\beta_1=\beta_2=90^\circ$, where the peak values of $\kappa_1$ and
$\kappa_2$ are used.
For small crossing angle $\theta_{\rm 12}=30^\circ$, the gain of in-plane modes around $\alpha_\perp \sim
0$ is always larger than the out-of-plane modes with
$|\alpha_\perp|<90^\circ$ irrespective of the beam polarization,
due to the rapidly dropping overlapping efficiency with increasing $\alpha_\perp$.
For larger crossing angles, however,
the relative importance of the out-of-plane modes with
respect to the in-plane modes are determined by the beam
polarization states 
to a greater extent.
As shown in Fig.~\ref{Fig:SP_GainL},
with consideration of the finite beam overlapping volume,
the gain of out-of-plane SP modes for large
crossing angle can still be larger than the in-plane modes when either beam I or beam
II deviates from s-polarization significantly.
Another noticeable thing is that for a given beam width $w_b$, the beam
overlapping region in the $(\mathbf{k}_{01},\mathbf{k}_{02})$-plane
is a rhombus with side length $w_b/\sin\theta_{12}$, which increases
with $\theta_{12}$ when $\theta_{12}\geq 90^\circ$. Together with
the fact that $\kappa_A^U$ increases with increasing $\theta_{12}$
as discussed above, $G_{\rm A,able}$ at $\theta_{12}=150^\circ$ is
much larger than other crossing angles, as shown in
Fig.~\ref{Fig:SP_GainL}. As a consequence, the larger obtuse
crossing angle is beneficial to SP mode amplification.
Finally, it is worth to point that the
plasma inhomogeneity can result in different phase matching length
along different spatial directions, imposing more limitation on the
available volume that can take part in the SP mode amplification.
Depending on the specific plasma condition, this can significantly
modify the relative importance of different SP modes.

\section{Discussion and summary}
\label{sec:conc}
In summary, based on a linear kinetic model for the shared plasma
modes of two overlapped laser beams, an analytic convective solution
is derived. From this solution, effects of wavelength difference,
crossing angle and polarization states of the two beams on the
collective SBS modes with shared IAW are discussed in details. A
small wavelength difference ($\sim $nm) is found to have negligible
effects on the SP modes except for very small crossing angle ($\sim
1^\circ$). For two beams with nearly equal wavelength, wavevectors
of the shared IAW of all possible collective SBS modes lie on a
circle in the bisecting plane between wavevectors of the two laser
beams. For a specified plasma condition, the wavelength of the
scattered waves decreases with increasing beam crossing angle
$\theta_{12}$ or increasing out-of-plane angle $\alpha_\perp$ of the
SP mode. However, the strength of the SP modes are subject to more
factors. Depending on the polarization states of the laser beams,
and the geometry and relative orientation of the gain volume
relative to the beam overlapping volume, the out-of-plane SP modes
can be important. Our results suggest that in a realistic simulation
for the collective SBS modes, all these factors must be properly
accounted for to obtain a reliable result.

In this work, uniform plasma conditions with zero flow velocity are
assumed for the illustrative analysis. However, it can be easily
extended to the non-zero flow velocity case, which would lead to a
wavelength shift of the scattered waves.
For practical ICF conditions, the plasma inhomogeneity, together with the practical
laser intensity distribution and overlapping pattern of the laser
beams, would complicate the situation significantly. Accurate
account of all these factors require simulations, for which this
work provides valuable theoretical references.
Furthermore, the
analytical convective solution presented here can help
construct the numerical solution by applying it over spatial regions of the grid size,
as done previously for single beam
LPIs~\cite{Strozzi2008RayBackScatter,Bezzerides1996SBSTwoIon}.

\section{acknowledgments}
This work was supported by the National Key R\&D Program of China
(Grant No.~2017YFA0403204), the Science Challenge Project (Grant
No.~TZ2016005), the National Natural Science Foundation of China
(Grant No.~11875093 and~11875091), and the Development Funds of CAEP
(Grant No.~CX20210040).

\bibliographystyle{unsrt}
\bibliography{citation}
\appendix
\section{Properties of ponderomotive response function}
\label{app:Gprop}
For the shared IAW modes of SBS,
one critical factor that appears in the gain coefficients is
\begin{equation}
  {k}_{a}^2\mathrm{Im}[\gamma_{\rm pm}]=k_{a}^2\mathrm{Im}[\frac{(1+\chi_I)\chi_e}{1+\chi_I+\chi_e}]
  \label{eq:Gfact}
\end{equation}
The peak value of ${k}_{a}^2\mathrm{Im}[\gamma_{\rm pm}]$
is achieved near the naturally resonant IAW modes where $\epsilon=1+\chi_I+\chi_e\approx 0$
and $\omega_a=k_ac_s$ with $c_s$ being the ion acoustic velocity.
Since $1+\chi_I+\chi_e\approx 0$, the peak value satisfies
\begin{equation}
  \max[{k}_{a}^2\mathrm{Im}[\gamma_{\rm pm}]] \approx \frac{k_{a}^2|\chi_e|^2}{(\partial \epsilon_r/\partial \omega) \nu_{a}},
  \label{eq:MaxGresp}
\end{equation}
which is inversely proportional to the damping rate ($\nu_a$) of
IAW. Here $\epsilon_r$ is the real part of $\epsilon$. It is found
that the peak value of ${k}_{a}^2\mathrm{Im}[\gamma_{\rm pm}]$
increases (weakly) with decreasing $k_{a}$, and the half-width of
${k}_{a}^2\mathrm{Im}[\gamma_{\rm pm}]$ which is proportional to the
damping is anti-correlated with the peak value of
${k}_{a}^2\mathrm{Im}[\gamma_{\rm pm}]$. This can be illustrated in
the fluid limit $v_{\rm the}\gg c_s\gg v_{\rm thI}$, where $v_{\rm
the}$ and $v_{\rm thI}$ are the electron and ion thermal velocity
respectively, and there are relatively simple analytical formulae
for $\chi_I$ and $\chi_e$. In the fluid limit,
\begin{equation}
   \max[{k}_{a}^2\mathrm{Im}[\gamma_{\rm pm}]]=\frac{\omega_{\rm pi}^2\omega_a}{2c_s^2\nu_{a}},
   \label{eq:pmfluid}
\end{equation}
where $\omega_{\rm pi}$ is the ion plasma frequency.
For He plasma with Maxwellian EEDF, the ion acoustic
velocity~\cite{Williams1995IAWTwoIon} is given by
\begin{equation}
  c_s=\sqrt{\frac{ZT_e}{m_i}} \sqrt{\frac{1}{1+k_a^2\lambda_{\rm De}^2}+\frac{3T_i}{ZT_e}},
\end{equation}
Using the approximation
$\nu_a=\mathrm{Im}[\epsilon]/(\partial \epsilon_r/\partial \omega)$, it can be obtained
\begin{equation}
  \frac{\nu_a}{\omega_a}=\sqrt{\frac{\pi}{8(1+k_a^2\lambda_{\rm De}^2)^3}}
  \left[(\frac{ZT_e}{T_i})^{3/2}\exp(-\frac{c_s^2}{2v_{\rm
  thI}^2})+\sqrt{\frac{Zm_e}{m_i}}\right].
\end{equation}
So with the decrease of $k_a$, $c_s$ (weakly) increases, and hence
the exponential part (corresponding to the tail of the ion energy
distribution function) of $\nu_a/\omega_a$ decreases. The change in
the exponential part is generally stronger, leading to decreasing
$\nu_a/\omega_a$ and $c_s^2\nu_a/\omega_a$ with decreasing $k_a$. As
a consequence, $\max[{k}_{a}^2\mathrm{Im}[\gamma_{\rm pm}]]$ as
given by Eq.~(\ref{eq:pmfluid}) increases (slowly) with decreasing
$k_a$.

\section{Convective solution for the SP modes of two crossing beams by the two-dimensional Laplace transformation}
\label{app:laplaceSP}
The two-dimensional Laplace transform of a function $f(x,y)$~\cite{Debnath1989MultiDimensionLaplace} is defined by
\begin{equation}
  \mathcal{L}_{\mathbf{q}}[f(x,y)]\equiv \int_0^\infty\int_0^\infty f(x,y)e^{-q^x x-q^y y}dxdy
\end{equation}
where $\mathbf{q}=(q^x,q^y)$,
and
$q^{x}$ and $q^{y}$ are the complex frequencies corresponding to $x_1$ and $x_2$, respectively.
The associated one-dimensional Laplace transforms along $x$-axis and $y$-axis are defined by
\begin{align}
  &\mathcal{L}_{q^x}[f(x)]\equiv \int_0^\infty f(x)e^{-q^x x}dx \\
  &\mathcal{L}_{q^y}[f(y)]\equiv \int_0^\infty f(y)e^{-q^y y}dy.
\end{align}
The Laplace transform changes differentiation to multiplication,
\begin{align}
  \mathcal{L}_{\mathbf{q}}[\partial_x f(x,y)]&=q^x\mathcal{L}_{\mathbf{q}}[f(x,y)]-\mathcal{L}_{q^y}[f(x=0,y)] \label{eq:Ldx} \\
  \mathcal{L}_{\mathbf{q}}[\partial_y f(x,y)]&=q^y \mathcal{L}_{\mathbf{q}}[f(x,y)]-\mathcal{L}_{q^x}[f(x,y=0)] \label{eq:Ldy}
\end{align}
When $\mathcal{L}_{\mathbf{q}}[f(x,y)]$ is known, $f(x,y)$ can be restored by the inverse Laplace transform,
\begin{equation}
f(x,y)=\mathcal{L}_{\mathbf{q}}^{-1}\{\mathcal{L}_{\mathbf{q}}[f(x,y)]\}.
  \label{eq:InvLq2d}
\end{equation}
The operation of $\mathcal{L}_{\mathbf{q}}^{-1}$ on the one-dimensional Laplace transforms yields
\begin{equation}
\begin{aligned}
  \mathcal{L}_{\mathbf{q}}^{-1}\{\mathcal{L}_{q^x}[f(x)]\}&=\delta(y)f(x) \\
  \mathcal{L}_{\mathbf{q}}^{-1}\{\mathcal{L}_{q^y}[f(y)]\}&=\delta(x)f(y)
\end{aligned}
  \label{eq:InvLq1d}
\end{equation}
The inverse Laplace transform changes convolution into multiplication,
\begin{equation}
  \mathcal{L}_{\mathbf{q}}^{-1}[G_1(\mathbf{q})G_2(\mathbf{q})]=\mathcal{L}_{\mathbf{q}}^{-1}[G_1(\mathbf{q})]*\mathcal{L}_{\mathbf{q}}^{-1}[G_2(\mathbf{q})]
  \label{eq:Lconv}
\end{equation}
where the symbol `*' denotes the convolution operation defined as
\begin{equation}
  (f*g)(x,y)=\int^x_0\int^y_0 f(u,v)g(x-u,y-v)dudv
  \label{eq:defconvolution}
\end{equation}

Using Eqs.~(\ref{eq:Ldx}-\ref{eq:Ldy}),
the two-dimensional Laplace transform of
Eqs.~(\ref{eq:scatter1}-\ref{eq:scatter2}) over the $(x_1,x_2)$-space can be written as
\begin{align}
& q^{x_1} \mathcal{L}_{\mathbf{q}}[a_{s_1}]-F_1 =\kappa_1(\mathcal{L}_{\mathbf{q}}[a_{s_1}]+r_a\mathcal{L}_{\mathbf{q}}[a_{s_2}]) \label{eq:scatterLap1} \\
  & q^{x_2} \mathcal{L}_{\mathbf{q}}[a_{s_2}]-F_2=\kappa_2(\mathcal{L}_{\mathbf{q}}[a_{s_2}]+\mathcal{L}_{\mathbf{q}}[a_{s_1}]/r_a)
  \label{eq:scatterLap2}
\end{align}
where
\begin{equation}
\begin{aligned}
& F_1\equiv \mathcal{L}_{q^{x_2}}[a_{s_1}({x_1}=0,{x_2})] \\
& F_2\equiv \mathcal{L}_{q^{x_1}}[a_{s_2}({x_1},{x_2}=0)]
\end{aligned}
  \label{eq:defF1F2}
\end{equation}
is determined by the boundary conditions specified by $a_{s_1}({x_1}=0,{x_2})$ and $a_{s_2}({x_1},{x_2}=0)$.
Solving for $\mathcal{L}_{\mathbf{q}}[a_{s_1}]$ and $\mathcal{L}_{\mathbf{q}}[a_{s_2}]$, we obtain
\begin{align}
  & \mathcal{L}_{\mathbf{q}}[a_{s_1}]=\frac{(q^{x_2}-\kappa_2)F_1+\kappa_1r_aF_2}{q^{x_1}q^{x_2}-\kappa_1q^{x_2}-\kappa_2q^{x_1}} \\
  & \mathcal{L}_{\mathbf{q}}[a_{s_2}]=\frac{(q^{x_1}-\kappa_1)F_2+\kappa_2F_1/r_a}{q^{x_1}q^{x_2}-\kappa_1q^{x_2}-\kappa_2q^{x_1}}.
  \label{eq:Lqa1s}
\end{align}
Then, $a_{s_1}$ and $a_{s_2}$ can be
restored from $\mathcal{L}_{\mathbf{q}}[a_{s_1}]$
and $\mathcal{L}_{\mathbf{q}}[a_{s_2}]$
by the inverse Laplace transform.
Using the identities (\ref{eq:InvLq2d}-\ref{eq:Lconv}), it can be obtained
\begin{equation}
  \begin{bmatrix} a_{s_1}({x_1},{x_2}) \\ a_{s_2}({x_1},{x_2}) \end{bmatrix}=
  \int_0^{x_2}\mathbf{G}_1(x_1,x_2-v)a_{s_1}({x_1}=0,v)dv+ \int_0^{x_1}\mathbf{G}_2(x_1-u,x_2)a_{s_2}(u,x_2=0)du
  \label{eq:SPas1as2App}
\end{equation}
where
\begin{equation}
  \mathbf{G}_1({x_1},{x_2})=\begin{bmatrix} G_{11} \\ G_{21} \end{bmatrix}=
  \begin{bmatrix}
 \kappa_2\kappa_1x_1e^{\kappa_1{x_1}+\kappa_2{x_2}}\frac{I_1[2\sqrt{\kappa_1\kappa_2{x_1}{x_2}}]}{\sqrt{\kappa_1\kappa_2{x_1}{x_2}}}\\
 (\kappa_2/r_a) e^{\kappa_1{x_1}+\kappa_2{x_2}}I_0[2\sqrt{\kappa_1\kappa_2{x_1}{x_2}}]
  \end{bmatrix}+
  \begin{bmatrix}
e^{\kappa_1{x_1}}\delta({x_2}) \\ 0
\end{bmatrix},~x_1\geq 0,x_2\geq 0
  \label{eq:RespG1App}
\end{equation}
and
\begin{equation}
  \mathbf{G}_2({x_1},{x_2})= \begin{bmatrix} G_{12} \\ G_{22} \end{bmatrix}=
  \begin{bmatrix}
  r_a\kappa_1 e^{\kappa_1{x_1}+\kappa_2{x_2}}I_0[2\sqrt{\kappa_1\kappa_2{x_1}{x_2}}] \\
  \kappa_1\kappa_2{x_2}e^{\kappa_1{x_1}+\kappa_2{x_2}}\frac{I_1[2\sqrt{\kappa_1\kappa_2{x_1}{x_2}}]}{\sqrt{\kappa_1\kappa_2{x_1}{x_2}}}
  \end{bmatrix}+
  \begin{bmatrix}
0 \\ e^{\kappa_2{x_2}}\delta({x_1})
\end{bmatrix},~x_1\geq 0,x_2\geq 0
  \label{eq:RespG2App}
\end{equation}

\section{The representative direction $\mathbf{n}_A$ and the representative gain coefficient $\kappa_A$}
\label{sec:SPmeasure}
The direction of $\mathbf{n}_A$ satisfies
$\mathbf{n}_A\parallel \nabla G_A$, where the
the asymptotic gain coefficient $G_A=e^{(\sqrt{\kappa_1x_1}+\sqrt{\kappa_2x_2})^2}$.
To obtain the direction of $\mathbf{n}_A$, it is convenient to rewrite $G_A$
in an orthogonal coordinate system.
Defining the $\tilde{x}_1$-axis along $\mathbf{n}_{s_1}$ and $\tilde{x}_2$-axis perpendicular to $\mathbf{n}_{s_1}$,
the transformation between ($x_1,x_2$) and ($\tilde{x}_1,\tilde{x}_2$) is given by
$x_1=(\tilde{x}_1\sin\theta_{12}^s-\tilde{x}_2\cos\theta_{12}^s)/\sin\theta_{12}^s$ and $x_2=\tilde{x}_2/\sin\theta_{12}^s$.
The condition $ \mathbf{n}_A\parallel \nabla G_A$ gives us
\begin{equation}
  \frac{{\partial G}/{\partial \tilde{x}_1}}{{\partial G}/{\partial \tilde{x}_1}}=\frac{\tilde{x}_1}{\tilde{x}_2},
  \label{eq:SPRefDire}
\end{equation}
After some manipulation,
the relation between
$x_r\equiv x_2/x_1$ and $\kappa_r\equiv {\kappa_2}/{\kappa_1}$ can be obtained,
\begin{equation}
  \frac{(x_r+\cos\theta_{12}^s)\sqrt{x_r}}{1+x_r\cos\theta_{12}^s}=\sqrt{\kappa_r}.
  \label{eq:SPRefDireCond}
\end{equation}
Correspondingly, the angle between $\mathbf{n}_A$ and $\mathbf{n}_{s_1}$ is
\begin{equation}
  \cos\theta_A=\frac{1+x_r\cos\theta_{12}^s}{\sqrt{1+x_r^2+2x_r\cos\theta_{12}^s}}.
  \label{eq:angRef}
\end{equation}
At the point
$\boldsymbol{x}=x_1(\mathbf{n}_{s_1}+x_r\mathbf{n}_{s_2})$ along the
direction $\mathbf{n}_A$,
$G_A=\exp[{(1+\sqrt{\kappa_rx_r})^2\kappa_1x_1}]$, and the required
2D gain volume is the parallelogram
$x_1(\eta_1\mathbf{n}_{s_1}+\eta_2 x_r \mathbf{n}_{s_2})$ ($0\leq
\eta_{1,2} \leq 1$). The two diagonals of the parallelogram have
lengths $x_1\sqrt{1+x_r^2\pm 2x_r\cos\theta_{12}^s}$. Using the root
mean square of the two diagonals as the typical size ($L_m$) of the
required gain volume, we have $L_m=x_1\sqrt{1+x_r^2}$, and the gain
coefficient $\kappa_A\equiv \ln G_A/L_m$ defined for this typical
size satisfies
\begin{equation}
  \mathcal{R}_{\rm A}[\kappa_r,\theta_{12}^s]\equiv\frac{\kappa_A}{\kappa_1+\kappa_2}=\frac{(1+\sqrt{{\kappa_rx_r}})^2}{(1+\kappa_r)\sqrt{1+x_r^2}}.
  \label{eq:GcoefR}
\end{equation}
The ratio $\mathcal{R}_{\rm A}$ is a function of $\kappa_r$ and
$\theta_{12}^s$, since $x_r$ in the RHS is a function of $\kappa_r$
and $\theta_{12}^s$ as specified by Eq.~(\ref{eq:SPRefDireCond}).
$\mathcal{R}_{\rm A}$ varies between $1$ and $\sqrt{2}$. The upper
bound of $\sqrt{2}$ appears with $\mathbf{n}_A$ along the bisector
of $\mathbf{n}_{s_1}$ and $\mathbf{n}_{s_2}$ when
$\kappa_1=\kappa_2$ and hence $x_1=x_2$, while the lower bound of
one can be approached when $\kappa_1 \ll \kappa_2$ or $\kappa_2 \ll \kappa_1$.
\section{The projection of the gain volume onto one direction $\mathbf{n}_r$ and perpendicular to $\mathbf{n}_r$}
\label{app:proj}
The parallelogram gain volume $V_{\rm amp}(\boldsymbol{x})=\eta_1x_1\mathbf{n}_{s_1}+\eta_2 x_2 \mathbf{n}_{s_2} (0\leq \eta_{1,2} \leq 1)$ is spanned by $\mathbf{n}_{s_1}$ and $\mathbf{n}_{s_2}$.
Denoting the projection of $\mathbf{n}_{s_\alpha}$ onto one direction $\mathbf{n}_r$ as $\mathbf{n}_{s_\alpha}^{\parallel \mathbf{n}_r}$,
and the projection of
$\mathbf{n}_{s_\alpha}$ onto the plane perpendicular to $\mathbf{n}_r$ as $\mathbf{n}_{s_\alpha}^{\perp \mathbf{n}_r}$,
then
\begin{equation}
\mathbf{n}_{s_\alpha}=
\mathbf{n}_{s_\alpha}^{\parallel \mathbf{n}_r}+ \mathbf{n}_{s_\alpha}^{\perp \mathbf{n}_r},
\end{equation}
yielding
\begin{equation}
  \begin{aligned}
  V_{\rm amp}& =V_{\rm amp}^{\parallel \mathbf{n}_r}+V_{\rm amp}^{\perp \mathbf{n}_r}, \\
  V_{\rm amp}^{\parallel \mathbf{n}_r}&\equiv x_1\eta_1\mathbf{n}_{s_1}^{\parallel \mathbf{n}_r}+x_2\eta_2\mathbf{n}_{s_2}^{\parallel \mathbf{n}_r} \\
V_{\rm amp}^{\perp \mathbf{n}_r}&\equiv x_1\eta_1\mathbf{n}_{s_1}^{\perp \mathbf{n}_r}+x_2\eta_2\mathbf{n}_{s_2}^{\perp \mathbf{n}_r}
  \end{aligned}
\end{equation}
It can be seen that
the longitude scale of the gain volume
defined as the length of $V_{\rm amp}^{\parallel \mathbf{n}_r}$ is
$\mathrm{Proj}_{\parallel \mathbf{n}_r}[V_{\rm amp}]=x_1|\mathbf{n}_{s_1}^{\parallel \mathbf{n}_r}|+x_2|\mathbf{n}_{s_2}^{\parallel \mathbf{n}_r}|$
for $\eta_{1,2}$ ranging from zero to one.
While $V_{\rm amp}^{\perp \mathbf{n}_r}$ is itself a parallelogram in the plane perpendicular to $n_{r}$,
so the transverse scale of the gain volume defined as the longest spatial scale of $V_{\rm amp}^{\perp \mathbf{n}_r}$
is the longer diagonal,
$\mathrm{Proj}_{\perp \mathbf{n}_r}[V_{\rm amp}]=\sqrt{x_1^2|\mathbf{n}_{s_1}^{\perp \mathbf{n}_r}|^2+x_2^2|\mathbf{n}_{s_2}^{\perp \mathbf{n}_r}|^2+2x_1x_2|\mathbf{n}_{s_1}^{\perp \mathbf{n}_r}\cdot\mathbf{n}_{s_2}^{\perp \mathbf{n}_r}|}$.
\end{document}